\documentclass[12pt]{article}
\pdfoutput=1

\usepackage[totalheight = 23cm, totalwidth = 17cm]{geometry}
\usepackage{amssymb,amsmath,amsfonts,amsbsy,graphicx}
\usepackage{color,bm,hyperref}
\PassOptionsToPackage{normalem}{ulem}
\usepackage{ulem}

\newcommand{\mpl}{m_{\rm Pl}}
\newcommand{\calH}{{\cal H}}

\newcommand{\calP}{{\cal P}}
\newcommand{\calR}{{\cal R}}

\begin{document}

\begin{titlepage}

\rightline{\footnotesize{APCTP-Pre2015-018}}

\begin{center}

\vskip 1.0cm

\textbf{\Huge Towards general patterns of features \\ in multi-field inflation
}

\vskip 1.0cm

\large{
Xian Gao$^{a}$ \, and \,
Jinn-Ouk Gong$^{b,c}$
}

\vskip 0.5cm

\small{\it 
$^{a}$Department of Physics, Tokyo Institute of Technology, Tokyo 152-8551, Japan
\\
$^{b}$Asia Pacific Center for Theoretical Physics, Pohang 790-784, Korea 
\\
$^{c}$Department of Physics, Postech, Pohang 790-784, Korea
}

\vskip 1.2cm

\end{center}

\begin{abstract}

We investigate the consequences of general curved trajectories in multi-field inflation. After setting up a completely general formalism using the mass basis, which naturally accommodates the notion of light and heavy modes, we study in detail the simple case of two successive turns in two-field system. We find the power spectrum of the curvature perturbation receives corrections that exhibit oscillatory features sinusoidal in the logarithm of the comoving wavenumber without slow-roll suppression. We show that this is because of the resonance of the heavy modes inside and outside the mass horizon.

\end{abstract}

\end{titlepage}

\setcounter{page}{0}
\newpage
\setcounter{page}{1}

\section{Introduction}

The recent cosmological observations by the Planck mission~\cite{Ade:2015xua} are mostly consistent with the predictions of the simplest single field inflation~\cite{Ade:2015lrj}: nearly scale invariant power spectrum of the curvature perturbation with almost perfect Gaussian statistics. Constructing a concrete model of inflation is, however, still an open question. For example, it is highly non-trivial to maintain the potential flat enough to support a long enough period of inflation, representatively illustrated by the so-called eta problem in supergravity~\cite{Copeland:1994vg}. Furthermore, many extensions of particle physics such as supersymmetric and/or string theory usually incorporate many scalar fields, which can in principle serve as the inflaton~\cite{Lyth:1998xn}. Thus, in more general context than the simplest model where a single, canonical inflaton minimally coupled to Einstein gravity drives inflation, the dynamics of inflation may well be more complicated. Indeed, we should take these general possibilities more seriously, as the new Planck data does not seem to favour the simple cases such as the $m^2\phi^2$ model~\cite{Ade:2015lrj}.


If the inflationary dynamics is more complicated, we can expect distictive observational signatures which may explain the anomalies in the power spectrum of the temperature fluctuations in the cosmic microwave background (CMB), such as the features in the correlation functions~\cite{Ade:2015lrj,featuresearch}. A good example is the case in which during multi-field inflation the trajectory is curved~\cite{Gordon:2000hv}: then the modes orthogonal to the trajectory can be excited by extracting the kinetic energy of the adiabatic component along the trajectory, equivalent to a single field theory with non-trivial speed of sound~\cite{heavy}. The resulting correlation functions of the curvature perturbation exhibit a localized oscillatory burst, and are correlated to each other~\cite{corrcorr}. The single field theory obtained by systematically integrating out the heavy modes is a very good effective description as long as the trajectory is adiabatic~\cite{singleeftvalidity}. When the trajectory is more ``suddenly'' curved~\cite{Gao:2012uq,Gao:2013ota,suddenturn} we have to take into account higher order corrections~\cite{Burgess:2012dz,Gong:2014rna}, and eventually the effective field theory (EFT) of single field inflation may enter a strongly coupled regime~\cite{strongcoupling}. See e.g.~\cite{Chluba:2015bqa} for a concise review.


While the theoretical detail of turning trajectories is developed and investigated in depth, in most literature the consequences are illustrated using the simplest example: when there is only a single turn during inflation, which gives rise to sinusoidal oscillations~\cite{heavy,Gao:2012uq,suddenturn,Gong:2014rna}. Meanwhile, other types of oscillations are known which originate from different theories. For example, logarithmic oscillations in the power spectrum~\cite{Jackson:2013mka} generically appear in models with non-Bunch-Davies initial conditions~\cite{nBD-logosc}, resonant oscillations of massive fields~\cite{reso-logosc} and axion monodromy model~\cite{Flauger:2014ana}. Thus, it is both important and interesting to take steps beyond the simplest case of a single turn to look for different patterns of oscillatory features. If, by considering more complicated curved trajectories than the one with a single turn, we can obtain other types of oscillations different from simple sinusoidal ones, this would indicate that broader -- or even possibly all --  classes of oscillations, available as analytic templates to search for parametrized features from the CMB data, can be generated in generic curved trajectories in multi-field inflation.


In this article, we take steps forward this goal by considering the next simplest case of two successive turns in the trajectory. Very interestingly, we find that even without any explicit background oscillations that are periodic in physical time, the power spectrum of the light mode can exhibit logarithmic oscillations after successive sudden turns of the inflationary trajectory. This relies in the fact that the massive mode behaves differently inside and outside of the ``mass horizon'', which oscillates periodically in comoving (physical) time inside (outside) the mass horizon. As a result, logarithmic oscillations arise if the two successive turns occurs when the massive mode is inside and outside of the mass horizon respectively. We may think of this new feature as a ``resonance'' between inside and outside of the mass horizon.

This article is outlined as follows. In Section~\ref{sec:formalism}, we set up the general formalism and introduce two different bases: the kinematic and mass bases. We explicitly discuss their meanings and relations in simple two-field case. In Section~\ref{sec:spectrum}, we consider a simple trajectory that experiences two successive turns and compute the corrections to the power spectrum of the curvature perturbation. We find that if the turns occur when the heavy modes of our interests are inside and outside of the mass horizon, the corrections exhibit oscillatory features sinusoidal in $\log{k}$. We then conclude in Section~\ref{sec:conc}.

\section{Dynamics along background trajectory}
\label{sec:formalism}

In this section, we first set up the general formalism which is valid for generic multi-field inflation. We then concentrate on two-field case to develop concrete picture. We mostly follow the notations of \cite{heavy,Gao:2012uq,Gao:2013ota,Gong:2002cx,Gao:2013zga}.

\subsection{General formalism}

We begin with the action
\begin{equation}
S = \int \! d^4x \sqrt{-g} \left[ \frac{\mpl^2}{2}R - \frac{1}{2}G_{ab}g^{\mu\nu}\partial_\mu\phi^a\partial_\nu\phi^b - V(\phi) \right] \, ,
\end{equation}
where $G_{ab}$ is a general field space metric with the latin indices $a$, $b$ and so on denoting general coordinates in the field space. Taking a flat FRW metric as our background
\begin{equation}
ds^2 = -dt^2 + a^2(t)\delta_{ij}dx^idx^j \, ,
\end{equation}
the $00$ and $0i$ components of the Einstein equation becomes
\begin{equation}
\begin{split}
3\mpl^2H^2 & = \frac{1}{2}\dot\phi_0^2 + V \, ,
\\
\dot{H} & = -\frac{\dot\phi_0^2}{2\mpl^2} \, ,
\end{split}
\end{equation}
where we have defined the rapidity of the evolution of the background field as
\begin{equation}
\label{eq:BG}
\dot\phi_0^2 \equiv G_{ab}\dot\phi_0^a\dot\phi_0^b \, .
\end{equation}
Thus the slow-roll parameter $\epsilon$ can be written as
\begin{equation}
\epsilon \equiv -\frac{\dot{H}}{H^2} = \frac{1}{2\mpl^2} \frac{\dot\phi_0^2}{H^2} \, .
\end{equation}
The background field equation follows from the standard Euler-Lagrange equation as
\begin{equation}
D_t\dot\phi_0^a + 3H\dot\phi_0^a + V^a = 0 \, ,
\end{equation}
where $V^a = G^{ab}V_{;b}$ with a semicolon being a covariant derivative with respect to the field space metric $G_{ab}$, and
\begin{equation}
\label{eq:bgeq}
D_t\dot\phi_0^a = \frac{D\dot\phi_0^a}{dt} \equiv \frac{d\dot\phi_0^a}{dt} + \Gamma^a_{bc}\dot\phi_0^b\dot\phi_0^c
\end{equation}
is a covariant time derivative acting on $\dot\phi_0^a$. The reason is because $\dot\phi_0^a$ is a vector that resides in the tangent space of the curved field space~\cite{Gong:2011uw}, thus we do need to take into account the contributions of the field space curvature.

We now turn to perturbations. Since we are interested in the situation where the background trajectory experiences successive turns that may lead to rapid oscillations, a relatively simple effective single field description calls for more care, although not impossible once sub-leading corrections in the EFT expansion are properly taken into account~\cite{Burgess:2012dz,Gong:2014rna}. Thus we do not integrate out the heavy degrees of freedom but compute the interactions with the light mode explicitly. That means we treat the degrees of freedom in the field contents as physical, and we do not give physical degrees of freedom to the spatial metric so that we choose the flat gauge where the spatial metric is unperturbed. Then the quadratic action is given by~\cite{Gong:2011uw,Langlois:2008mn}
\begin{equation}
S_2 = \int \! d^4x \frac{a^3}{2} \left( G_{ab}D_tQ^aD_tQ^b - \frac{\delta^{ij}}{a^2}G_{ab}\partial_iQ^a\partial_jQ^b - C_{ab}Q^aQ^b \right) \, ,
\end{equation}
where $Q^a = \phi^a - \phi_0^a$ that is equivalent to the field fluctuation $\delta\phi^a$ at linear order~\cite{Burgess:2012dz,Gong:2011uw} and 
\begin{equation}
C_{ab} \equiv V_{;ab} - R_{acdb}\dot\phi_0^c\dot\phi_0^d + (3-\epsilon)\dot\phi_{0a}\dot\phi_{0b} + \frac{1}{H} \left( \dot\phi_{0a}V_{;b} + \dot\phi_{0b}V_{;a} \right) \, .
\end{equation}

Now we rewrite the action in terms of the perturbations orthogonal to each other by introducing a complete set of vielbeins $e_a^I = e_a^I(t)$ which maps the general, arbitrary basis denoted by $a$ into a local orthogonal frame denoted by $I$ as
\begin{equation}
e^I_ae^J_bG^{ab} = \delta^{IJ} \quad \text{and} \quad e^I_ae^J_b\delta_{IJ} = G_{ab} \, .
\end{equation}
The field fluctuation $Q^a$ can be transformed into the one in the orthogonal frame by incorporating $e^I_a$ as
\begin{equation}
Q^I = e^I_aQ^a \, .
\end{equation}
Then, introducing $u^I \equiv aQ^I$ and moving to the conformal time $d\eta = dt/a$, we obtain
\begin{equation}
\label{eq:S2}
S_2 = \int \! d\eta d^3x \frac{1}{2} \left[ \delta_{IJ} \left( \frac{du^I}{d\eta}\frac{du^J}{d\eta} + 2\frac{du^I}{d\eta}Z^J{}_Ku^K + Z^I{}_KZ^J{}_Lu^Ku^L - \delta^{ij}\partial_iu^I\partial_ju^J \right) - a^2 M_{IJ} u^Iu^J \right] \, ,
\end{equation}
where $M_{IJ} \equiv C_{IJ} - H^2(2-\epsilon)\delta_{IJ}$ with $C_{IJ} \equiv e^a_Ie^b_JC_{ab}$, and $Z^I{}_J \equiv e^I_aD_\eta e^a_J$. We may choose whatever frame we like for the $IJ$ basis as long as it is orthogonal. A physically important one is the so-called ``kinematic basis''~\cite{Gordon:2000hv,heavy}, which is set along and perpendicular to the field trajectory. The unit tangent vector $T^a$ is defined by 
\begin{equation}
T^a \equiv \frac{\dot\phi_0^a}{\dot\phi_0} \, .
\end{equation}
The normal vector $N^a$ which satisfies $G_{ab}T^aN^b = 0$ is naturally proportional to the derivative of $T^a$, i.e. $D_tT^a \propto N^a$. We {\em define} the proportionality parameter as the angular velocity $\dot\theta$ of the trajectory,
\begin{equation}
D_tT^a \equiv \dot\theta N^a \, ,
\end{equation}
which means 
\begin{equation}
\dot\theta = N_aD_tT^a \, .
\end{equation}
Note that $\dot\theta$ here is defined with the opposite sign from~\cite{heavy}.

\subsection{Two-field case: kinematic and mass bases}

\begin{figure}[t]
\begin{center}
 \includegraphics[width = 0.6\textwidth]{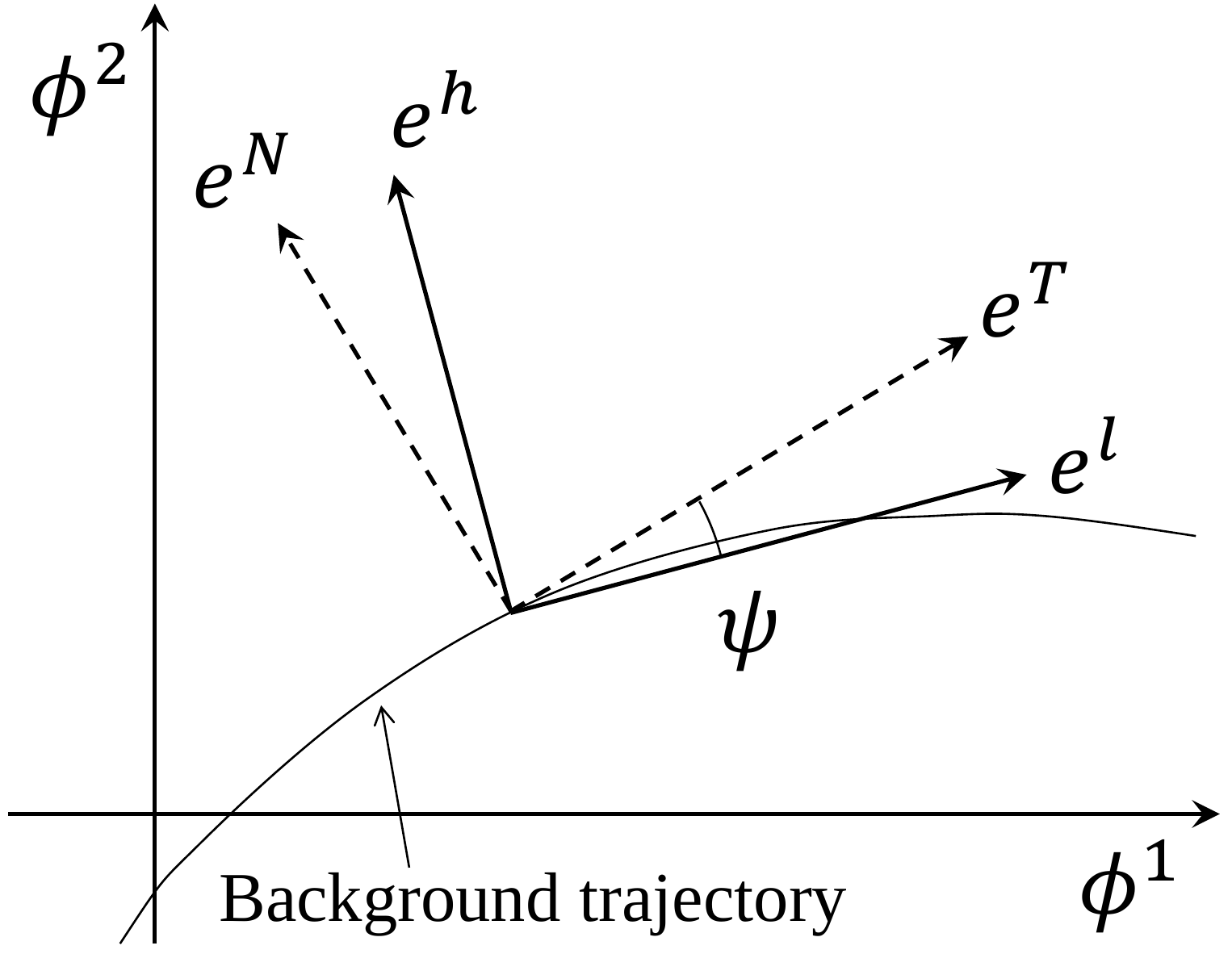}
\end{center}
\caption{A schematic plot showing the relation between the kinematic basis $\{e^T,e^N\}$and mass one $\{e^l,e^h\}$ related by rotation by an angle $\psi$. The angular velocity of the kinematic basis is $\dot\theta$ while that of the mass basis is $\dot\vartheta$, so that $\theta$, $\vartheta$ and $\psi$ are related by $\theta = \vartheta+\psi$, although for $\theta$ and $\vartheta$ only their rates of change are important.}
\label{fig:rotation}
\end{figure}

We now consider the case in which there are two fields. This is the simplest and thus most intuitive case with multiple number of fields, yet captures many important aspects of multi-field inflation. When we talk about ``heavy'' or ``light'' degrees of freedom, they do not necessarily coincide with the tangent and normal components to the trajectory. That is, the kinematic basis is not the set of eigenvectors of the mass matrix $V_{;IJ}$. For two-field case, one of the vielbeins corresponds to the tangent vector $T^a$ and the other to the normal vector $N^a$,
\begin{equation}
e^a_1 = T^a \quad \text{and} \quad e^a_2 = N^a \, .
\end{equation}
Then the mass matrix in the kinematic basis is given by
\begin{equation}
\label{eq:VIJ}
V_{;IJ} = 
\begin{pmatrix}
V_{TT} & V_{TN} 
\\
V_{TN} & V_{NN}
\end{pmatrix}
\, ,
\end{equation}
where in general $V_{TN} \neq 0$ for a curved trajectory. Thus, in terms of ``heavy'' and ``light'' degrees of freedom, it is most convenient to adopt another set of basis which makes the mass matrix diagonal. We may call this as the ``mass basis''~\cite{Gao:2012uq,Gao:2013ota}\footnote{Note the terminology used in this work is different from that in \cite{Gao:2012uq,Gao:2013ota}. In the current work, we refer to $V_{;IJ}$ as the ``mass matrix'', and thus the ``mass basis'' in which $V_{;IJ}$ is diagonalized in this work corresponds to the ``potential basis'' in \cite{Gao:2012uq,Gao:2013ota}.}. Then the eigenvalues of the mass basis corresponds to the light and heavy masses along the trajectory. Note that the situation is different from the change-of-basis around the bottom of the potential trough~\cite{Burgess:2012dz} where we do not need any kinematic information but everything is determined by geometry. Explicit diagonalization of \eqref{eq:VIJ} gives two eigenvalues,
\begin{equation}
\label{eq:VllVhh}
\lambda_\pm = \frac{1}{2} \left[ V_{NN}+V_{TT} \pm \left( V_{NN}-V_{TT} \right) \sqrt{1+\beta^2} \right] \quad \text{where} \quad \beta \equiv \frac{2V_{TN}}{V_{NN}-V_{TT}} \, ,
\end{equation}
so that $\lambda_-$ ($\lambda_+$) corresponds to $V_{ll}$ ($V_{hh}$). The corresponding eigenvectors transformed from the kinematic basis are then 
\begin{equation}
\begin{split}
e_l^a & = T^a\cos\psi - N^a\sin\psi = 
\begin{pmatrix}
\cos\psi
\\
-\sin\psi
\end{pmatrix}
\, ,
\\
e_h^a & = T^a\sin\psi + N^a\cos\psi =
\begin{pmatrix}
\sin\psi
\\
\cos\psi
\end{pmatrix}
\, ,
\end{split}
\end{equation}
with
\begin{equation}
\cos\psi \equiv \frac{1+\sqrt{1+\beta^2}}{\sqrt{2\left(1+\beta^2+\sqrt{1+\beta^2}\right)}}
\quad \text{and} \quad
\sin\psi \equiv \frac{\beta}{\sqrt{2\left(1+\beta^2+\sqrt{1+\beta^2}\right)}} \, .
\end{equation}
The meaning is very clear: the change-of-basis matrix $P$ constructed from $e_l^a$ on the first and $e_h^a$ on the second column respectively is
\begin{equation}
\label{eq:k-to-m}
P = \begin{pmatrix}
\cos\psi & \sin\psi
\\
-\sin\psi & \cos\psi
\end{pmatrix}
\, ,
\end{equation}
so that basically we rotate the orthogonal basis $\{T^a,N^a\}$ by $\psi$. This situation is depicted in Figure~\ref{fig:rotation}. Note that $\theta$ means the rotation angle from the generic basis $\{\phi^1,\phi^2\}$ to the kinematic one $\{T^a,N^a\}$. Thus the value of $\theta$ is not important since $\phi^a$, viz. the field space coordinates can be set totally arbitrary, but only the rate of its change $\dot\theta$, the ``angular velocity'' of the trajectory, is important. However, $\psi$ means the misalignment between the kinematic and mass bases and thus not only its rate of change but also its value are important to describe the dynamics along the trajectory: a non-zero $\psi$ would lead to a trajectory with oscillations caused by the misalignment between the kinematic and mass bases, and these oscillations will be damped and eventually disappear as the two bases coincide.

Finally, $M_{IJ}$ in \eqref{eq:S2} in the mass basis is 
\begin{align}
\label{eq:coupling-Mll}
M_{ll}^2 & = V_{ll} - 2\epsilon H^2 \left[ \left( 3-\epsilon+\frac{\dot\epsilon}{H\epsilon} \right)\cos^2\psi - \frac{\dot\theta}{H}\sin(2\psi) + \mpl^2\mathbb{R}\sin^2\psi \right] - H^2(2-\epsilon) \, ,
\\
\label{eq:coupling-Mlh}
M_{lh}^2 & = 2\epsilon H^2 \left[ -\frac{1}{2}\left( 3-\epsilon+\frac{\dot\epsilon}{H\epsilon} \right)\sin(2\psi) - \frac{\dot\theta}{H}\cos(2\psi) + \frac{1}{2}\mpl^2\mathbb{R}\sin(2\psi) \right] \, ,
\\
\label{eq:coupling-Mhh}
M_{hh}^2 & = V_{hh} - 2\epsilon H^2 \left[ \left( 3-\epsilon+\frac{\dot\epsilon}{H\epsilon} \right)\sin^2\psi + \frac{\dot\theta}{H}\sin(2\psi) + \mpl^2\mathbb{R}\cos^2\psi \right] - H^2(2-\epsilon) \, ,
\end{align}
where
\begin{equation}
\mathbb{R} \equiv R_{abcd}T^aN^bT^cN^d
\end{equation}
is the curvature of the field space. Then the quadratic action \eqref{eq:S2} becomes explicitly
\begin{align}
\label{eq:S2-2}
S_2 & = \int \! d\eta d^3x \frac{1}{2} \left[ {u_l'}^2 - (\nabla u_l)^2 - \left( a^2M_{ll}^2-{\vartheta'}^2 \right)u_l^2 \right.
\nonumber\\
& \qquad\qquad\qquad + {u_h'}^2 - (\nabla u_h)^2 - \left( a^2M_{hh}^2-{\vartheta'}^2 \right)u_h^2
\nonumber\\
& \qquad\qquad\qquad - 4\vartheta'u_l'u_h - 2 \left( a^2M_{lh}^2 + \vartheta'' \right) u_lu_h \bigg] \, ,
\end{align}
where $\vartheta'$ denotes the angular velocity of the mass basis. The first and second lines denote the free terms of the light and heavy modes respectively, while the third line is the interaction between them. Note that we have partially integrated some terms to remove time derivatives from the heavy modes. Also note that $\theta$, $\vartheta$ and $\psi$ are related by $\theta = \vartheta + \psi$.

\section{Changes in the power spectrum}
\label{sec:spectrum}

Having found the quadratic action in the previous section, we can proceed to find the power spectrum of the comoving curvature perturbation $\calR$, which is given by the standard relation
\begin{equation}
\mathcal{R} \equiv  -\frac{H}{\dot{\phi}_0} T_{a} Q^a = -\frac{H}{\dot{\phi}_0} \frac{1}{a} \left( u_l \cos\psi + u_h \sin\psi \right) \, .
\end{equation}
Before and long after the turns, the mass basis coincides with the kinematic basis, i.e. $\psi\rightarrow 0$, and thus $\mathcal{R} = -Hu_l/(\dot\phi_0a)$.
For our purpose, we may simply concentrate on the power spectrum of the light mode $u_l$. From below, we assume for simplicity that $V_{ll}$ and $V_{hh}$ are approximately constant and the field space is flat. Extensions to more general cases is however straightforward.

From \eqref{eq:S2-2}, we can straightly find the corresponding Hamiltonian in the interaction picture as
\begin{equation}
H(\eta) = \int \frac{d^3k}{(2\pi)^3} \left( \calH_0 + \calH_\text{turn} + \calH_\text{nsr} \right) \, ,
\end{equation}
where\footnote{Note that $\calH_\text{turn}$ is apparently different from the corresponding equation (2.37) in \cite{Gao:2013ota}, since here we performed an integration by parts.}
\begin{align}
\calH_0 & = \frac{1}{2} \left( {u_l'}^2 + k^2u_l^2 - \frac{z_0''}{z_0} u_l^2 \right) + \frac{1}{2} \left[ {u_h'}^2 + k^2u_h^2 + \left( a_0^2V_{hh} - \frac{a_0''}{a_0} \right) u_h^2 \right] \, ,
\\
\calH_\text{turn} & = -\frac{1}{2}{\vartheta'}^2u_l^2 + 2\vartheta'u_l'u_h + \vartheta''u_lu_h + \frac{3}{2}{\vartheta'}^2u_h^2 \, ,
\\
\calH_\text{nsr} & = \frac{1}{2} \left( a^2M_{ll}^2 + \frac{z_0''}{z_0} \right)u_l^2 + a^2M_{lh}^2u_lu_h + \frac{1}{2}\left( a^2M_{hh}^2 - a_0^2V_{hh} + \frac{a_0''}{a_0} \right)u_h^2 \, ,
\end{align}
with $z\equiv a\phi_0'/\calH$. These describe respectively a free Hamiltonian in a slow-roll background denoted by a subscript 0, the interactions due to the turning trajectory with explicit $\vartheta$ dependence and the departure from the slow-roll background. The free solutions for the mode functions $u_l$ and $u_h$ are then respectively found as
\begin{align}
\label{eq:ul_free}
u_l(\eta,k) & = \frac{1}{\sqrt{2k}} \left( 1 - \frac{i}{k\eta} \right) e^{-ik\eta} \, ,
\\
\label{eq:uh}
u_h(\eta,k) & = \frac{\sqrt{\pi}}{2} \exp\left( \frac{-\pi}{2}\nu + i\frac{\pi}{4} \right) \sqrt{-\eta}H_{i\nu}^{(1)}(-k\eta) \, ,
\end{align}
where $H_\nu^{(1)}$ is the first kind of the Hankel function with the index being given by
\begin{equation}
\nu \equiv \sqrt{\frac{V_{hh}}{H_0^2} - \frac{9}{4}} \, .
\end{equation}
Then, according to the standard in-in formalism, the corrections to the power spectrum is given by
\begin{equation}
\frac{\Delta\calP_\calR}{\calP_\calR} = \Delta_\text{turn1} + \Delta_\text{turn2} + \Delta_\text{nsr1} + \Delta_\text{nsr2} \, ,
\end{equation}
where $\calP_\calR$ is the free power spectrum constructed from \eqref{eq:ul_free}, and
\begin{align}
\label{eq:dPturn1}
\Delta_\text{turn1} & = -2\Im \left[ \int_{-\infty}^0 d\eta {\vartheta'}^2(\eta)u_l^2(\eta) \right] \, ,
\\
\label{eq:dPturn2}
\Delta_\text{turn2} & = 2 \left| \int_{-\infty}^0 d\eta \left[ \vartheta''(\eta)u_l(\eta) + 2\vartheta'(\eta)u_l'(\eta) \right] u_h(\eta) \right|^2
\nonumber\\
& \quad + 4\Re \left[ \int_{-\infty}^0 d\eta_1 \left[ \vartheta''(\eta_1)u_l(\eta_1) + 2\vartheta'(\eta_1)u_l'(\eta_1) \right] u_h^*(\eta_1) \right.
\nonumber\\
& \qquad\qquad \times \left. \int_{-\infty}^{\eta_1} d\eta_2 \left[ \vartheta''(\eta_2)u_l(\eta_2) + 2\vartheta'(\eta_2)u_l'(\eta_2) \right] u_h(\eta_2) \right] \, ,
\\
\label{eq:dPnsr1}
\Delta_\text{nsr1} & = 2\Im \left[ \int_{-\infty}^0 d\eta \left( a^2M_{ll}^2 + \frac{z_0''}{z_0} \right) u_l^2(\eta) \right] \, ,
\\
\label{eq:dPnsr2}
\Delta_\text{nsr2} & = 2 \left| \int_{-\infty}^0 d\eta a^2M_{lh}^2 u_l(\eta)u_h(\eta) \right|^2
+ 4\Re \left[ \int_{-\infty}^0 d\eta_1 a^2M_{lh}^2 u_l(\eta_1)u_h^*(\eta_1) \int_{-\infty}^{\eta_1} d\eta_2 a^2M_{lh}^2 u_l(\eta_2)u_h(\eta_2) \right] \, .
\end{align}

We are interested in evaluating these corrections when the field trajectory is more complicated than experiencing a single turn. In practice, $\vartheta$ can be parametrized by the form
\begin{equation}
\vartheta = \sum_i \alpha_iS_i(\eta) \, ,
\end{equation}
where $\alpha_i$ is the turning angle during a single turning process labeled by $i$, and $S_i(\eta)$ is some smooth function of time interpolating between 0 and 1. Let us consider, for the next simplest case, two instantaneous turns, with
\begin{equation}
\label{eq:2turns}
\vartheta'(\eta) = \alpha_1\delta(\eta-\eta_1) + \alpha_2\delta(\eta-\eta_2) \, ,
\end{equation}
where $\alpha_1$ and $\alpha_2$ are two turning angles at the corresponding moments $\eta_1$ and $\eta_2$ respectively, with 
\begin{equation}
\eta_1 < \eta_2 < 0 \, .
\end{equation}

\subsection{Background solutions}

To evaluate first \eqref{eq:dPnsr1} and \eqref{eq:dPnsr2}, we need to find the background solutions away from the otherwise standard slow-roll ones. The necessary background equations to be solved are obtained by projecting \eqref{eq:bgeq} onto $T^a$ and $N^a$. We can write those equations in several different forms as we like. Here, we choose to work with the following ones~\cite{Gao:2012uq,Gao:2013ota,Gao:2013zga}:
\begin{align}
\label{eq:bgepsilon}
\frac{\epsilon''}{2\epsilon} + \left[ \calH(1-2\epsilon)-\frac{\epsilon'}{4\epsilon} \right] \frac{\epsilon'}{\epsilon} + 2\calH^2(\epsilon-3)\epsilon & = {\theta'}^2 - a^2V_{TT} \, ,
\\
\label{eq:bgtheta}
\theta'' + 2\left[ \calH(1-\epsilon) + \frac{\epsilon'}{2\epsilon} \right] \theta' & = -a^2V_{TN} \, .
\end{align}
The reason why we write the background equations in terms of $\epsilon$ and $\theta$ is because these variables explicitly appear in the couplings for the perturbations, as can be read from \eqref{eq:coupling-Mll}, \eqref{eq:coupling-Mlh}, \eqref{eq:coupling-Mhh} and \eqref{eq:S2-2}. Thus once we find the solutions of these equations we can directly use them to evaluate \eqref{eq:dPnsr1} and \eqref{eq:dPnsr2} without further manipulations.

The detailed steps to solve these equations are already illustrated in~\cite{Gao:2012uq,Gao:2013ota}, so we only briefly sketch the basic strategy and give the results with the ansatz for the trajectory of our current interest \eqref{eq:2turns}. Assuming $\psi \ll 1$, \eqref{eq:bgtheta} can be approximated by
\begin{equation}
\psi'' - \frac{2}{\eta}\psi' + \frac{1}{\eta^2} \left( \tilde\nu^2 + \frac{9}{4} \right)\psi = -\vartheta'' + \frac{2}{\eta}\vartheta' \, , 
\end{equation}
where
\begin{equation}
\tilde\nu \equiv \sqrt{\frac{V_{hh}-V_{ll}}{H_0^2} - \frac{9}{4}} \, .
\end{equation}
Then, for a given $\vartheta'$, $\psi$ can be easily found by using the retarded Green's function as
\begin{equation}
\psi(\eta) = -(-\eta)^{3/2}\sec\varphi \int_{-\infty}^\eta d\tau (-\tau)^{-3/2} \cos \left[ \tilde\nu\log \left( \frac{\eta}{\tau} \right) + \varphi \right] \vartheta'(\tau) \, ,
\end{equation}
with the phase $\varphi$ being given by
\begin{equation}
\varphi \equiv \tan^{-1}\left( \frac{3}{2\tilde\nu} \right) \, .
\end{equation}
For \eqref{eq:2turns}, we can find
\begin{equation}
\label{eq:solpsi}
\psi(\eta) = -\sec\varphi \sum_{i=1}^2 \Theta(\eta-\eta_i) \alpha_i \left( \frac{\eta}{\eta_i} \right)^{3/2} \cos \left[ \tilde\nu\log \left( \frac{\eta}{\eta_i} \right) + \varphi \right] \, ,
\end{equation}
where $\Theta(x)$ is the Heaviside step function which equals to unity if $x>0$ otherwise is vanishing.

We can solve \eqref{eq:bgepsilon} by noting that the non-trivial part of $\epsilon$ is induced by the departure from usual slow-roll by \eqref{eq:2turns}. First we split $\epsilon$ into two parts $\epsilon=\epsilon_0+\Delta\epsilon$, and approximate $\calH$ and $a$ by their smooth values. This is justified by the fact that with $\Delta\epsilon$ oscillating with a frequency $\tilde\nu\gg1$, the deviations from the slow-roll values of $\calH$ and $a$ are suppressed by $\tilde\nu$. Then, \eqref{eq:bgepsilon} implies a linear equation for $\Delta\epsilon$ as
\begin{align}
\label{eq:Deltaepsilon}
& \Delta\epsilon'' - \frac{2\left(1-2\epsilon_0\right)}{\eta} \Delta\epsilon' - \frac{12\epsilon_0}{\eta^2} \Delta\epsilon
\nonumber\\
& = - \frac{2\epsilon_0}{\eta^2} \tilde\nu^2 \sec^3\varphi \sum_{i,j=1}^2 \Theta\left(\eta-\eta_i\right) \Theta\left(\eta-\eta_j\right) \alpha_i\alpha_j \left( \frac{-\eta}{\sqrt{\eta_i\eta_j}} \right)^3 \cos\left[ 2\tilde{\nu}\log \left( \frac{-\eta}{\sqrt{\eta_i\eta_j}} \right)+\varphi \right] \, .
\end{align}
By noting that $\Theta\left(\eta-\eta_1\right) \Theta\left(\eta-\eta_2\right) = \Theta\left(\eta-\eta_2\right)$ since $\eta_1<\eta_2$, the solution to \eqref{eq:Deltaepsilon} is given by
\begin{equation}
\label{eq:solDe}
\Delta\epsilon = \frac{1}{2}\epsilon_0 \sec^2\varphi \sum_{i,j=1}^2 \Theta(\eta-\eta_i)\Theta(\eta-\eta_j) \alpha_i\alpha_j \left( \frac{-\eta}{\sqrt{\eta_i\eta_j}} \right)^3 \cos \left[ 2\tilde\nu\log \left( \frac{-\eta}{\sqrt{\eta_i\eta_j}} \right) + 2\varphi \right] \, , 
\end{equation}
where we have suppressed the terms without any oscillatory contributions. Note that this solution can be compared with (3.15) in \cite{Gao:2013ota}.

\subsection{Contributions from non-slow-roll background}

We now evaluate the contributions from the departure from the smooth slow-roll background, \eqref{eq:dPnsr1}. This is studied in detail in~\cite{Gao:2013ota} so we are satisfied here to describe important steps only. For our purpose, first we need to determine the coefficient of the self coupling $a^2M_{ll}^2+z_0''/z_0$. Expanding $M_{ll}^2$ in small $\psi$ and $\Delta\epsilon$ and then using the solutions for them given by \eqref{eq:solpsi} and \eqref{eq:solDe} respectively, we can show that to leading order in this expansion the only remaining term is given by\footnote{Interestingly, this is also consistent with (3.16) in~\cite{Gao:2013ota}, where it was \emph{assumed} that the oscillatory part in $a^2M_{ll}^2+z_0''/z_0$ mainly comes from $a''/a$, i.e. the last term in $M_{ll}^2$ given by \eqref{eq:coupling-Mll}. Here we show that this is actually exact, since the expanded terms from the second term in \eqref{eq:coupling-Mll} exactly cancel out up to leading order in $\Delta\epsilon$ and $\psi^2$.}
\begin{equation}
a^2M_{ll}^2 + \frac{z_0''}{z_0} = H_0^2\Delta\epsilon \, .
\end{equation}
With this coupling, we can recast \eqref{eq:dPnsr1} as
\begin{align}
\label{eq:dPnsr1-2}
\Delta_\text{nsr1} = \frac{1}{4}\epsilon_0 \sec^2\varphi \, \Im\left\{ \sum_{i,j=1}^2 \alpha_i\alpha_j \int_0^{\min\left\{ x_i,x_j\right\}} \frac{dx}{x^2} \left( 1+\frac{i}{x} \right)^2 \left( \frac{x}{\sqrt{x_ix_j}} \right)^3 \left[e^{i\Phi_{+}\left(x,\sqrt{x_{i}x_{j}}\right)}+e^{i\Phi_{-}\left(x,\sqrt{x_{i}x_{j}}\right)}\right] \right\} \, , 
\end{align}
where we have introduced $x \equiv -k\eta$, $x_1\equiv -k\eta_1$, $x_2\equiv -k\eta_2$ and
\begin{equation}
\Phi_{\pm}\left(x,\sqrt{x_{i}x_{j}}\right) \equiv 2x \pm \left[ 2\tilde{\nu} \log\left( \frac{x}{\sqrt{x_{i}x_{j}}} \right) + 2\varphi \right] \, .
\end{equation}
Keep in mind that since $\eta_1 < \eta_2 <0$, we thus have $\min \{x_1, x_2\} = x_2$. Although the integral \eqref{eq:dPnsr1-2} can be evaluated analytically, to have an intuitive understanding, here we use the stationary phase approximation\footnote{The relevant formula is well-known: when $\lambda\gg1$,
\[
\int_{a}^{b}\mathrm{d}x\,g\left(x\right)e^{i\lambda\Phi\left(x\right)}\approx g\left(c\right)\sqrt{\frac{2\pi}{\lambda\left|\Phi''\left(c\right)\right|}}\exp\left[i\lambda\,\Phi\left(c\right)+\frac{i\pi}{4}\mathrm{sgn}\left\{ \Phi''\left(c\right)\right\} \right] 
\]
for a real-valued function $\Phi\left(x\right)$.}. In our case, since $x>0$, only $\Phi_{-}\left(x,\sqrt{x_{i}x_{j}}\right)$ reaches its stationary point at $x_{-} = \tilde{\nu}$, and thus the corresponding integral contributes provided that $\min\left\{ x_{i},x_{j}\right\} > \tilde{\nu} $.
After some manipulations, for $\tilde{\nu}\gg 1$, finally we have
\begin{align}
\label{eq:dPnsr1-f}
\Delta_\text{nsr1}(k) & \approx \frac{\sqrt{\pi}}{4} \epsilon_0 \tilde{\nu}^{3/2} \, \sum_{i,j=1}^2 \alpha_{i}\alpha_{j} \Theta\left( \frac{k}{\tilde{\nu}\,\max\left\{ k_{i},k_{j}\right\} }-1 \right) 
\nonumber\\
& \qquad\qquad\qquad \times \left( \frac{\sqrt{k_{i}k_{j}}}{k} \right)^{3} \sin \left[ 2\tilde{\nu}\log \left( \frac{k}{\tilde{\nu}\sqrt{k_{i}k_{j}}} \right) + 2\tilde{\nu} + \frac{\pi}{4} \right] \, ,
\end{align}	
with $k_i = -1/\eta_i$. Again, since $\eta_1 < \eta_2<0$, we actually have $\max \{k_1,k_2\} = k_2$. As a comparison, note that \eqref{eq:dPnsr1-f} recovers (3.26) in~\cite{Gao:2013ota} when there is a single turn. It is also important to note that \eqref{eq:dPnsr1-f} is slow-roll suppressed.

To evaluate \eqref{eq:dPnsr2}, we first note that from \eqref{eq:coupling-Mlh}
\begin{equation}
\theta'(\eta) = -\tilde\nu \sqrt{-\eta} \sec^2\varphi \int_{-\infty}^\eta dx (-x)^{-3/2} \sin\left[ \tilde\nu \log\left( \frac{\eta}{x} \right) \right] \vartheta'(x) \, .
\end{equation}
Then from \eqref{eq:coupling-Mlh} we can find that \eqref{eq:dPnsr2} is suppressed exponentially (first term) and by power-law (second term)~\cite{Chen:2012ge}, so that \eqref{eq:dPnsr2} provides smaller corrections compared with \eqref{eq:dPnsr1-f}.

\subsection{Contributions from turns}

Now we evaluate the contributions to the power spectrum due to the turning of the trajectory, which are the in-in formalism integrals $\Delta_\text{turn1}$ and $\Delta_\text{turn2}$  given in \eqref{eq:dPturn1} and \eqref{eq:dPturn2} respectively. Apparently, $\Delta_\text{turn1}$ is divergent in the case of instantaneous turns since the factor ${\vartheta'}^2$ enters in the integrand\footnote{This can also be seen when the turn has a finite time duration, i.e. a finite strength, which is modeled by a Gaussian function $d\vartheta(t)/dt = \Delta\theta\mu\exp \left( -\mu^2t^2/2 \right)/\sqrt{2\pi}$ as in~\cite{Gao:2012uq,Gao:2013ota}. In~\cite{Gao:2012uq}, the integral corresponding to $\Delta_\text{turn1}$ is analytically evaluated, which is shown to be proportional to $\mu$ [see (5.23), (5.28) and (5.32) there] and thus diverges when the turn becomes infinitely sharp, $\mu\rightarrow \infty$. However, as we show here in this article, this divergence is artificial as it will be exactly cancelled when the full contributions are taken into account.}. This divergence, however, is artificial as $\Delta_\text{turn2}$ contains exactly the same divergence with an opposite sign. In fact, it can be shown that as long as the turning process has a finite time duration, i.e.
\begin{equation}
\vartheta'(\eta=-\infty) = \vartheta'(\eta=0) = 0 \, ,
\end{equation}
the total contribution $\Delta_\text{turn} = \Delta_\text{turn1} + \Delta_\text{turn2}$ can be equivalently recast to be
\begin{align}
\label{eq:dPturn}
\Delta_\text{turn} & = 2 \left| \int_{-\infty}^{0} d\eta\,\vartheta'\left(\eta\right) \left[ u_{l}'\left(\eta,k\right) u_{h}\left(\eta,k\right) - u_{l}\left(\eta,k\right) u_{h}'\left(\eta,k\right) \right] \right|^{2}
\nonumber\\
& \quad + 4\, \Re \bigg\{ \int_{-\infty}^{0} d\eta_{1} \, \vartheta'\left(\eta_{1}\right) \left[ u_{l}'\left(\eta_{1},k\right) u_{h}^{\ast}\left(\eta_{1},k\right) - u_{l}\left(\eta_{1},k\right) u_{h}'^{\ast}\left(\eta_{1},k\right) \right]
\nonumber \\
& \qquad\qquad \times \int_{-\infty}^{\eta_{1}} d\eta_{2} \, \vartheta'\left(\eta_{2}\right) \left[ u_{l}'\left(\eta_{2},k\right) u_{h}\left(\eta_{2},k\right) - u_{l}\left(\eta_{2},k\right) u_{h}'\left(\eta_{2},k\right) \right] \bigg\} \, .
\end{align}
It is now manifest that this expression is always finite even in the case of instantaneous turns. We emphasize that \eqref{eq:dPturn} is exact and general.

We now use \eqref{eq:dPturn} to evaluate the contributions due to the double instantaneous turns \eqref{eq:2turns}. Straightforward manipulations yield
\begin{equation}\label{I_turn_fin}
\Delta_\text{turn} = \sum_{i,j=1}^2 \mathcal{I}_{ij} \, ,
\end{equation}
with
\begin{align}
\label{I_turn_ij}
\mathcal{I}_{ij} & \equiv 2\alpha_{i}\alpha_{j} \, \Re \Big\{ u_{h}(\eta_{i},k) u_{h}^{\ast}(\eta_{j},k) \, \Pi_{-}(\eta_{i},k)
\nonumber \\
& \qquad\qquad\quad \times \left[ u_{l}(\eta_{i},k) u_{l}^{\ast}(\eta_{j},k) \, \Pi_{-}^{\ast}(\eta_{j},k) + 2\Theta(\eta_{j}-\eta_{i}) u_{l}(\eta_{i},k) u_{l}(\eta_{j},k) \, \Pi_{+}(\eta_{j},k) \right] \Big\} \, , 
\\
\label{Pipm}
\Pi_{\pm}(\eta,k) & \equiv \frac{\partial}{\partial\eta} \log \left| \frac{u_{l}(\eta,k)}{u_{h}(\eta,k)} \right| - \frac{i}{2} \left( \frac{1}{\left|u_{l}(\eta,k)\right|^{2}} \pm \frac{1}{\left|u_{h}(\eta,k)\right|^{2}} \right) \, .
\end{align}
Note that $\Pi_{\pm}$ contains no oscillatory factors since only the moduli of the mode functions enter. As a check, let us consider the case where there is only a single turn with $\alpha\equiv\alpha_{\ast}$ and turning time $\eta_{\ast}$. Then \eqref{I_turn_fin} reduces to
\begin{equation}
\mathcal{I}_{11} = 2\alpha_\ast^{2} \left| u_{h}\left(\eta_{\ast},k\right) \right|^{2} \Big\{ \left| u_{l}\left(\eta_{\ast},k\right) \right|^{2} \left| \Pi_{-}\left(\eta_{\ast},k\right) \right|^{2} + 2\Theta\left(0\right) \, \Re \left[ u_{l}^{2}\left(\eta_{\ast},k\right) \Pi_{+}\left(\eta_{\ast},k\right) \Pi_{-}\left(\eta_{\ast},k\right) \right] \Big\} \, ,
\end{equation}	
where $\Theta(0)$ is some numerical constant between $0$ and $1$. It is thus clear that the heavy mode function contributes through the modulus, so that there is only sinusoidal oscillations arising from the second term inside the curly brackets. Such oscillations are studied in detail in~\cite{Gao:2012uq}.

\begin{figure}[t]
\begin{center}
 \includegraphics[width=0.6\textwidth]{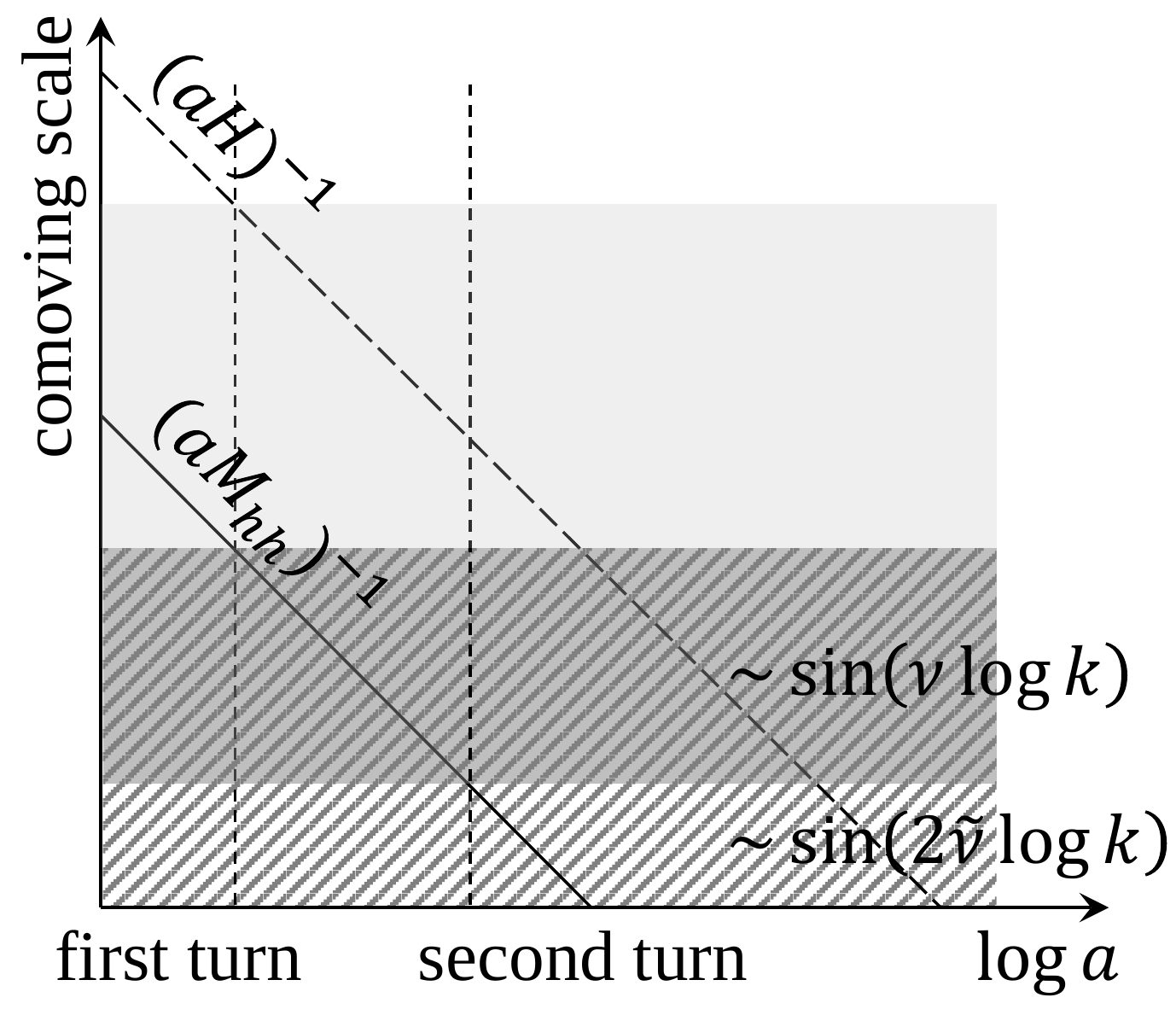}
\end{center}
\caption{A schematic plot illustrating the respective scales of oscillatory features. If the wavelength of the given mode is outside of the mass horizon when both turns occur (but still inside the respective Hubble radius), there is only oscillations sinusoidal in $k$ (light shade region). If the wavelength of the given mode is inside the mass horizon when the first turn occurs while outside of the mass horizon when the second turn occur, there will be oscillations sinusoidal in $\log k$, with frequency $\nu$ (dark shade region). This is the new feature identified in this work. While modes with wavelength shorter than the mass horizon generally exhibits oscillations sinusoidal in $\ln k$, but with frequency $2 \tilde{\nu}$ (region with slanting lines).}
\label{fig:scales}
\end{figure}

Now we turn to \eqref{eq:2turns}. When there are two turns, due to the above argument, both $\mathcal{I}_{11}$ and  $\mathcal{I}_{22}$ contain only oscillations sinusoidal in $k$. However, $\mathcal{I}_{12}$ and $\mathcal{I}_{21}$ can contain oscillations sinusoidal in $\log k$. To see this, first note we have 
\begin{align}
\label{I_turn_1221}
\mathcal{I}_{12} + \mathcal{I}_{21} & = 4 \alpha_{1}\alpha_{2} \, \Re \Big\{ u_{h}(\eta_{1},k) u_{h}^{\ast}(\eta_{2},k) \Pi_{-}(\eta_{1},k)
\nonumber \\
&  \qquad\qquad\quad \times \left[ u_{l}(\eta_{1},k) u_{l}^{\ast}(\eta_{2},k) \Pi_{-}^{\ast}(\eta_{2},k) + u_{l}(\eta_{1},k) u_{l}(\eta_{2},k) \Pi_{+}(\eta_{2},k) \right]\Big\} \, .
\end{align}
Now comes the crucial point. Unlike the light mode, whose behavior is sensitive only to the Hubble scale, the heavy mode \eqref{eq:uh} behaves differently inside and outside of the ``mass horizon'', of which the length scale is $({\nu} a H)^{-1}$ with ${\nu}$ defined in \eqref{eq:uh}. To be precise, depending on whether it is inside or outside the mass horizon, $u_h$ can be approximated differently as~\cite{Gao:2012uq}
\begin{equation}
u_{h}\left(\eta,k\right) \approx
\begin{cases}
\dfrac{e^{-ik\eta}}{\sqrt{2k}} \left( 1 + i\dfrac{4{\nu}^{2}+1}{8k\eta} \right) 
& \text{for } -k\eta\gg{\nu} \, ,
\\
\dfrac{e^{-i\nu\left[\log\left(2{\nu}\right)-1\right]}}{\sqrt{2{\nu}}} \sqrt{-\eta} e^{i{\nu}\log\left(-k\eta\right)} & \text{for } -k\eta\lesssim{\nu} \, .
\end{cases}
\end{equation}
Thus, if the two turning time satisfy
\begin{equation}
\label{time}
-k\eta_{1}\gg{\nu} \quad \text{and} \quad -k\eta_{2}\lesssim {\nu} \, , 
\end{equation}
there will be oscillations sinusoidal in $\log k$ hidden in the factor 
\begin{equation}
u_h\left(\eta_{1},k\right) u_h^{\ast}\left(\eta_{2},k\right) \sim e^{-ik\eta_{1}+i\nu\log\left(-k\eta_{2}\right)}
\end{equation} 
in the first line of \eqref{I_turn_1221}. \eqref{time} implies when the first turn happens at $\eta=\eta_1$, the heavy mode is deep inside the mass horizon and oscillates as a free plane wave. Meanwhile when the second turn happens at $\eta=\eta_2$, the heavy mode has been outside of the mass horizon (although might or might not be still inside the Hubble radius) and thus oscillates periodically in physical time $t \sim \log (-\eta)$. Thus the arising of such oscillations may be thought of as ``resonance'' between inside and outside of the mass horizon. This situation is depicted in Figure~\ref{fig:scales}. If we further assume $-k\eta_2 \ll 1$, straightforward manipulations yield
\begin{equation}
\label{I_turn_1221_f}
\Delta_\text{turn} \supset \mathcal{I}_{12} + \mathcal{I}_{21} \approx \frac{1}{2}\alpha_{1}\alpha_{2} \, \nu^{9/2} \frac{k_{1}^{2}k_2^{1/2}}{k^{5/2}} \,\cos\left\{ \nu\log\left(\frac{k}{k_{2}}\right) + 2\frac{k}{k_{1}} - \nu \left[ \log\left(2\nu\right)-1 \right] \right\} \, .
\end{equation}
This is one of the main results in this article. Thus we discover a new mechanism to generate logarithmic osicllations in the power spectrum, which is alternative to the usual resonance due to the background oscillations. \eqref{I_turn_1221_f} implies that even without any explicit background oscillations, the final power spectrum may exhibit logarithmic osicllations after successive turns. Compared with \eqref{eq:dPnsr1-f} which also exhibits logarithmic oscillations, $\Delta_\text{turn}$ is not slow-roll suppressed, and thus will be the dominant contribution to the logarithmic osicllations in our system.

\section{Conclusion}
\label{sec:conc}

In this article, we have started from the observation that while the general formalism for multi-field inflation with curved trajectories is established, the phenomenology of such a curved trajectory is only illustrated using the simplest case with a single turn along an otherwise straight trajectory. A single turn gives rise to sinusoidal oscillations in the correlation functions of the curvature perturbation, but there are other types of oscillations which arise from different origins. Thus it is both very interesting and important to take further steps to exploit possible patterns of oscillations by considering more general trajectories in multi-field inflation. We have made first few steps towards this direction.

First we have summarized the general formalism of perturbations by introducing a general orthogonal basis on the background trajectory. A particular attention has been paid to the two-field system and the mass basis, which is not necessarily equivalent to the kinematic basis but in general is misaligned by $\psi$. The mass basis is particularly useful when we consider the notion of ``light'' and ``heavy'' modes, which are most naturally described in the mass basis because the mass matrix is diagonal.

Having set up the general two-field system, then we have considered the next simplest case of two successive turns. The corrections to the power spectrum of the curvature perturbation can be classified into two categories, those from the departure from the usual slow-roll background and those from the turning of the trajectory. The latter contributions have explicit couplings in terms of the angular velocity of the mass basis. The corrections from the non-slow-roll background exhibit logarithmic oscillations, which however arise even there is a single turn and are slow-roll suppressed. The main contributions to the logarithmic oscillations come from the direct coupling due to turns. The logarithmic oscillations appear when the heavy modes resonate inside and outside the mass horizon at the moment of two successive turns.

We would like to emphasize that in this article we have only considered the simplest trajectory beyond a single turn. Nevertheless we have found logarithmic oscillations in the power spectrum which are supposed to have other theoretical origins. We thus expect that in more generic trajectories different oscillatory features can arise, and the possibility of these patterns is vast. It would be worth going beyond the simple case study in this article to investigate more general trajectories and the consequences as well as the non-linear phenomenology such as non-Gaussianity.

\subsection*{Acknowledgments}

XG would like to thank the Asia Pacific Center for Theoretical Physics for hospitality during his visit where this work was initiated. 
JG is grateful to the Tokyo Institute of Technology for hospitality while this work was under progress.
XG is supported by JSPS Grant-in-Aid for Scientific Research No. 25287054 and 26610062.
JG acknowledges the Max-Planck-Gesellschaft, the Korea Ministry of Education, Science and Technology, Gyeongsangbuk-Do and Pohang City for the support of the Independent Junior Research Group at the Asia Pacific Center for Theoretical Physics. JG is also supported by a Starting Grant through the Basic Science Research Program of the National Research Foundation of Korea (2013R1A1A1006701).

\end{document}